\documentclass[iop]{emulateapj}
\usepackage{amssymb,amsmath,epsfig,times,natbib,longtable,color,textcomp}

\newcommand{\chandra}{\textit{Chandra}}
\newcommand{\nustar}{\textit{NuSTAR}}
\newcommand{\suzaku}{{\it Suzaku}}

\begin{document}

\lefthead{\nustar\ observations of the hard state in Cyg~X-1}
\righthead{Parker et al.}

\title[NuSTAR observations of the hard state in Cyg~X-1]{NuSTAR and Suzaku observations of the hard state in Cygnus~X-1: locating the inner accretion disk}

\author{
M. L. Parker\altaffilmark{1},
J. A. Tomsick\altaffilmark{2},
J. M. Miller\altaffilmark{3},
K. Yamaoka\altaffilmark{4},
A. Lohfink\altaffilmark{1},
M. Nowak\altaffilmark{5},
A. C. Fabian\altaffilmark{1},
W. N. Alston\altaffilmark{1},
S. E. Boggs\altaffilmark{2},                                                                        
F. E. Christensen\altaffilmark{6},                                                                  
W. W. Craig\altaffilmark{2,7},                                                                      
F. F\"urst\altaffilmark{8},                                                                         
P. Gandhi\altaffilmark{9},                                                                          
B. W. Grefenstette\altaffilmark{8},                                                                 
V. Grinberg\altaffilmark{5},                                                                        
C. J. Hailey\altaffilmark{10},                                                                      
F. A. Harrison\altaffilmark{8},                                                                     
E. Kara\altaffilmark{1},                                                                            
A. L. King\altaffilmark{11},                                                                        
D. Stern\altaffilmark{12},                                                                          
D. J. Walton\altaffilmark{8,12},                                                                    
J. Wilms\altaffilmark{13},                                                                          
and W. W. Zhang\altaffilmark{14}}   
                                                                
\altaffiltext{1}{Institute of Astronomy, Madingley Road, Cambridge, CB3 0HA}                        
\altaffiltext{2}{Space Sciences Laboratory, University of California, Berkeley, 7 Gauss Way, Berkeley, CA 94720-7450, USA}                                                                              
\altaffiltext{3}{Department of Astronomy, University of Michigan, 1085 South University Avenue, West Hall 311, Ann Arbor, MI 48109-1042, USA}                                                           
\altaffiltext{4}{Solar-Terrestrial Environment Laboratory, Department of Particles and Astronomy, Nagoya University, Furocho, Chikusa-ku, Nagoya, Aichi 464-8601, Japan}                                
\altaffiltext{5}{Massachusetts Institute of Technology, Kavli Institute for Astrophysics, Cambridge, MA 02139, USA}                                                                                     
\altaffiltext{6}{Danish Technical University, DK-2800 Lyngby, Denmark}                              
\altaffiltext{7}{Lawrence Livermore National Laboratory, Livermore, CA, USA}                        
\altaffiltext{8}{California Institute of Technology, 1200 East California Boulevard, Pasadena, CA 91125, USA}                                                                                           
\altaffiltext{9}{School of Physics \& Astronomy, University of Southampton, Highfield, Southampton SO17 1BJ, UK}
\altaffiltext{10}{Columbia University, New York, NY 10027, USA}
\altaffiltext{11}{Department of Physics, Stanford University, 382 Via Pueblo Mall, Stanford, CA 94305, USA}
\altaffiltext{12}{Jet Propulsion Laboratory, California Institute of Technology, 4800 Oak Grove Drive, Pasadena, CA 91109, USA}
\altaffiltext{13}{Dr Karl Remeis-Observatory and Erlangen Centre for Astroparticle Physics, Sternwartstr. 7, 96049 Bamberg, Germany}
\altaffiltext{14}{NASA Goddard Space Flight Center, Greenbelt, MD 20771, USA}

\begin{abstract}
We present simultaneous \emph{Nuclear Spectroscopic Telescope Array} (\nustar\ ) and \suzaku\ observations of the X-ray binary Cygnus~X-1 in the hard state. This is the first time this state has been observed in Cyg~X-1 with \nustar , which enables us to study the reflection and broad-band spectra in unprecedented detail. 
We confirm that the iron line cannot be fit with a combination of narrow lines and absorption features, and instead requires a relativistically blurred profile in combination with a narrow line and absorption from the companion wind. We use the reflection models of \citet{Garcia14} to simultaneously measure the black hole spin, disk inner radius, and coronal height in a self-consistent manner.
Detailed fits to the iron line profile indicate a high level of relativistic blurring, indicative of reflection from the inner accretion disk. We find a high spin, a small inner disk radius, and a low source height, and rule out truncation to greater than three gravitational radii at the $3\sigma$ confidence level.
In addition, we find that the line profile has not changed greatly in the switch from soft to hard states, and that the differences are consistent with changes in the underlying reflection spectrum rather than the relativistic blurring.
We find that the blurring parameters are consistent when fitting either just the iron line or the entire broad-band spectrum, which is well modelled with a Comptonized continuum plus reflection model. 
\end{abstract}


\section{Introduction}

The hard state in X-ray binaries (XRBs) is characterised by a spectrum with only weak thermal emission from the disk, a hard Comptonized component, and reflection from the accretion disk. It is generally agreed that the change between soft and hard states must be due to a change in the geometry of the accreting system, but there is still controversy over the exact nature of the change. This disagreement hinges on whether the accretion disk is significantly truncated in the hard state. It is generally agreed that the disk is truncated at a radius of $\sim100$ gravitational radii ($~r_\textrm{G}$) as binaries go into quiescence \citep[e.g.][]{Tomsick09}, but at what stage truncation begins as the accretion rate drops remains uncertain. In the disk truncation model \citep{Esin97}, as the accretion rate drops the classical thin disk becomes unstable and is truncated. This radius is typically greater than $\sim$10--20~$~r_\textrm{G}$, and the inner disk is replaced with a radiatively inefficient advection dominated accretion flow \citep[ADAF;][]{Narayan94, Narayan96, Blandford99}. This behaviour has been invoked to explain the transition from soft to hard states. However, some authors claim instead that the disk extends to small radii in the hard state while the accretion rate is higher than $\sim1$ per cent of the Eddington rate, as measured using the profile of the relativistic iron line \citep[e.g.][]{Miller06a,Miller06b,Miller15,Reis10}.

There are two main ways of probing the inner edge of the thin disk, both of which rely on measuring the relativistic effects that occur close to the black hole. The first involves modelling the relativistic blurring of the Fe K$\alpha$ line emitted when coronal X-rays are reflected from the inner accretion disk \citep{Fabian89}. This technique has been used extensively to measure black hole spins, both in XRBs and active galactic nuclei \citep[AGN; see review by][]{Reynolds13}. The second method, known as continuum fitting, relies on the black body spectrum of the disk itself \citep[see review by][]{McClintock14}. There are difficulties with using either method to look at XRBs in the hard state, as both reflection and disk components are much weaker in this state than in the soft state. In this work we will focus on modelling the iron line, as the disk black body lies outside the energy range covered by \nustar .
The expected behaviour of the reflection spectrum, if the disk truncation model is correct, is for the iron K line to become narrower as the disk truncates, since the relativistic blurring becomes less extreme. While this is difficult to measure due to the relative weakness of the reflection component in the hard state, truncation has been claimed in some sources at relatively high accretion rates \citep[e.g.][]{Plant15}. This also provides a natural explanation for the lower reflection fraction in the hard state, as far less of the Comptonized emission encounters the thin disk if it is truncated, assuming a compact corona.

Further confusing the issue is the degeneracy between source height and inner radius \citep{Dauser13,Fabian14}. For simple coronal models where the source is compact and on-axis \citep[the `lamp-post' geometry; see][for discussion of more complex geometries]{Wilkins12}, the narrow lines predicted by truncation can instead be interpreted as an increase in the distance between the source and the disk, which leads to less extreme light-bending and more illumination of the outer disk.

Cygnus X-1 (Cyg~X-1) was the first confirmed black hole, when its mass was determined to be too great for it to be a neutron star \citep[e.g.][]{Webster72,Tananbaum72,Bolton72,Gies86}. It was one of the first X-ray sources discovered, using Geiger counters aboard a sounding rocket \citep{Bowyer65}, and has been extensively studied since then. Several authors have investigated the inner disk in the hard state in this source. \citet{Done99} used a combination of \emph{ASCA}, \emph{EXOSAT} and \emph{Ginga} data to investigate the hard state, finding significant relativistic blurring, consistent with a moderately truncated disk. \citet{Young01} use data from the same three instruments and argue in favour of a non-truncated disk around a Schwarzchild black hole. \citet{Frontera01} compared the soft and hard states using \emph{BeppoSAX} data, finding inner radii of $<10r_\textrm{G}$ for the soft state and $10_{-4}^{+5}r_\textrm{G}$ in the hard state. \citet{Reis10} found an inner radius consistent with the ISCO of a Schwarzchild  black hole using \suzaku\ data. \citet{Miller12} found that the inner edge of the accretion disk in Cyg~X-1 was consistent with being located at the innermost stable circular orbit (ISCO) using reflection modelling of 20 \suzaku\ observations of the hard state, and \citet{Fabian12cygx1} found that the relativistic reflection required high spin, fitting an average observation from the same dataset. Measurements of the spin of Cyg~X-1 in the soft state using continuum fitting and and relativistic reflection have also returned high spin values \citep{Gou11,Tomsick14}, and the same has been found during transition \citep{Duro11}. Therefore, a systematic difference in the spin measurement between states would be indicative of truncation of the disk.

Here we present \nustar\ and \suzaku\ observations of Cyg~X-1 in the hard state. The soft state of Cyg~X-1 was previously observed with \nustar\ and \suzaku\ \citep{Tomsick14}, but this is the first time that the hard state of Cyg~X-1 has been seen with the ground-breaking sensitivity and bandpass of \nustar .
The launch of the \nustar\ X-ray telescope \citep{Harrison13} has revolutionized the study of the X-ray spectra of XRBs, as \nustar 's high sensitivity, broad energy range, and triggered read-out allow for spectra of unprecedented quality \citep{Miller13,Miller13Serpens,Fuerst13,Fuerst14,Natalucci14,Tendulkar14}. In this paper we focus on modelling the broad band spectrum of the Cyg~X-1 hard state. Timing and variability analysis will appear in future work. In particular, we are interested in measuring the spin and inner radius of the disk, so that we can establish whether the disk is truncated, and if so by how much.

\section{Observations and Data Reduction}
\label{section_datareduction}

Cyg~X-1 was observed for $\sim35$~ks with \nustar\ and $\sim107$~ks with \suzaku . 
The details of all observations are given in Table~\ref{table_exposures}. In this section we separately describe the observations and data reduction for each instrument.

\begin{table}
\centering
\caption{Times, exposures and count rates for the two \nustar\ detectors (FPMA and FPMB) and three \suzaku\ detectors (XIS1, PIN and GSO). Exposure times are after filtering for background flares. Count rates are for the energy bands used for spectral fitting: 3--79~keV for \nustar\ ; 1.2--1.7 and 2.5--9~keV for XIS1; 20--38 and 43--70~keV for the PIN; and 60--300~keV for the GSO.}
\label{table_exposures}
\begin{tabular}{l l c c r}
\hline
\hline
Instrument &Start Time &On source & count rate\\
&(UT)&exposure (ks)& (s$^{-1}$)\\
\hline
 FPMA & 2014 May 20th, 09hr & 34.4 & 205\\
FPMB & " & 35.3 & 189\\
 XIS1 & 2014 May 20th, 07hr& 12 & 134\\
  PIN & 2014 May 19th, 07hr& 107 & 9.6\\
 GSO & "& 107& 8.4\\
\hline
\end{tabular}
\end{table}

\subsection{NuSTAR}
The \nustar\ data (ObsID 30001011007) for the two focal plane modules (FPMs) were reduced using the latest versions of the \nustar\ data analysis software (NuSTARDAS v1.4.1) and CALDB (20140414). Cleaned event files were produced using the \textsc{nupipeline} routine, and spectral products with \textsc{nuproducts}. We used a 150$^{\prime\prime}$ circular extraction region for the source spectrum, and a 100$^{\prime\prime}$ region for the background, taken from an area of the detector not strongly contaminated by source counts (i.e. the furthest away part of the field from Cyg~X-1). The source flux is at least an order of magnitude higher than that of the background over the whole band, so variations in the background with position or off-axis angle will not affect our results. We fit the \nustar\ data over the whole energy range of 3--79~keV.
The \nustar\ spectra are binned to oversample the spectral resolution by a factor of 3, and to a minimum signal to noise ratio of 50 after background subtraction. At low energies in both detectors the oversampling restriction dominates (below $\sim20$~keV), as the total number of counts is extremely large (the FPMA spectrum alone contains almost 5 million counts).

\subsection{Suzaku}

For {\em Suzaku}, we reduced the X-ray Imaging Spectrometer \citep[XIS;][]{Koyama07} data from ObsID 409049010 using HEASOFT v6.16 and the 2014 July 1 version of the calibration files. The XIS coverage began on 2014 May 20, 6.7 h UT and ended on 2014 May 22 7.4 h UT. Although there are normally three operational XIS units, only XIS1 was on during this observation due to limitations on the available power during the late stages of the mission. We produced two separate XIS1 event lists using {\ttfamily aepipeline}, one of the full observation and one only including the first half of the observation (from the start time given above to 2014 May 21, 9.1 h UT), corresponding to the time of the {\em NuSTAR} coverage.  We ran {\ttfamily aeattcor2} and {\ttfamily xiscoord} to update the photon positions. Due to the brightness of Cyg~X-1 XIS1 was operated in 1/4 window mode, which reduces pile-up at the expense of reduced exposure time, and used the 0.3~s burst option. Even with the window mode, we used {\ttfamily pileest} and found $>$4\% pile-up within $1^{\prime}$ of the center of the point spread function.  For making the XIS1 spectrum, we used an annular extraction region with an inner radius of $1^{\prime}$ and an outer radius of $4^{\prime}$. After this region is excluded, the photon averaged pile-up fraction is 1.6\% and the pixel averaged fraction is 0.7\%.  A background spectrum was taken from a rectangular region near the edge of the field of view. A portion of the source extraction region includes inactive regions of the detector, and we accounted for this by adjusting the BACKSCAL keyword in the source spectral file. After creating the source and background spectral files, we used {\ttfamily xisrmfgen} and {\ttfamily xissimarfgen} to produce the response matrix. In the spectral fitting, we included the 1.2--1.7\,keV and 2.5--9.0\,keV ranges. There are known calibration uncertainties associated with the Si K-edge in the 1.7--2.1\,keV band and we extend the excluded range up to 2.5~keV because of the presence of narrow features potentially due to calibration errors. We bin the spectra to oversample the spectral resolution by a factor of 3, and to a minimum signal to noise ratio of 30 after background subtraction. The oversampling restriction dominates below $\sim8$ keV, so the full instrumental resolution is maintained over the iron line band. We do not add any systematic error to the XIS data.

The hard X-ray detector \citep[HXD;][]{Takahashi07} consists of Silicon PIN diodes covering an energy range of 12--70 keV and GSO scintillators covering an energy range of 40--600 keV. Cyg X-1 was observed at the XIS nominal position with a net HXD exposure of 107.0 ks, and detected up to 400 keV by GSO. The HXD data was analyzed in the standard manner using the perl script {\tt hxdpinxbpi} and {\tt hxdgsoxbpi} for PIN and GSO respectively. The modeled non X-ray background (NXB) was taken from public ftp sites\footnote{ftp://legacy.gsfc.nasa.gov/suzaku/data/background/pinnxb\_ver2.2\_tuned/ and ftp://legacy.gsfc.nasa.gov/suzaku/data/background/gsonxb\_ver2.6/}, and cosmic X-ray background (CXB) was also subtracted based on previous HEAO observations \citep{Gruber99} for PIN. A 1~per~cent systematic error was added for each spectral bin. In the spectral fitting, we used the energy responses: ae\_hxd\_pinxinome11\_20110601.rsp for PIN and ae\_hxd\_gsoxinom\_20100524.rsp with an additional correction file (ae\_hxd\_gsoxinom\_crab\_20100526.arf) for GSO. We fit the PIN data from 20--70~keV, excluding lower energies to avoid thermal noise and the 38--43~keV band to avoid a known calibration feature, and the GSO data from 60--300~keV. We extracted spectra for both HXD detectors corresponding to the intervals simultaneous and non-simultaneous with the \nustar\ exposure. We find no significant differences in the spectral shape for either detector between the two intervals, so we use the full exposures in all fits.

\section{Results}
\label{section_results}

We split the spectral fitting into three sections: in \S\ref{section_continuum} we fit simple continuum models to the broad-band spectrum to establish the presence and energy of the high-energy cut-off without modelling the reflection spectrum; in \S\ref{section_ironline} we fit the iron line profile in detail, to measure the relativistic blurring without complications from other regions of the spectrum; finally, in \S\ref{section_fullspectrum} we fit the full broad-band spectrum, including the reflection, to calculate our final best fit model. We use \textsc{Xspec} \citep{Arnaud96} version 12.8.2a for all fitting. 
All errors are quoted at 1 standard deviation unless otherwise stated.

\subsection{Broad-band continuum fitting}
\label{section_continuum}

To establish a baseline continuum model from which to measure the reflection properties we initially fit the \nustar\ FPMA and FPMB and \suzaku\ XIS, PIN and GSO data simultaneously. For this section only, we include a 5 per cent systematic error to each spectral bin for all datasets, to ensure that the model is being fit to the broad-band spectrum rather than being dominated by small but high signal-to-noise features in the \nustar\ spectra. We stress that the $\chi^2$ values presented in this section should therefore not be taken as absolute indications of the fit quality, only relative - without the systematic error, the reduced $\chi^2$ values are on the order of 35--40. In all fits we include interstellar absorption, modelled with \textsc{TBabs} \citep[using the abundances of][]{Wilms00}, fixed at the value of $6\times 10^{21}$~cm$^{-2}$ from \citet{Tomsick14}. We use only the simultaneous XIS1 for the broad-band fits (\S~\ref{section_continuum} and \ref{section_fullspectrum}), as this ensures a consistent power law index $\Gamma$ between the XIS and FPM spectra. The full XIS1 spectrum gives a significantly different value of $\Gamma$, making it much more difficult to simultaneously fit the XIS and \nustar\ spectra. We include a constant normalization offset between the instruments, to allow for differences in flux calibration.

\begin{figure*}
\centering
\includegraphics[width=\textwidth]{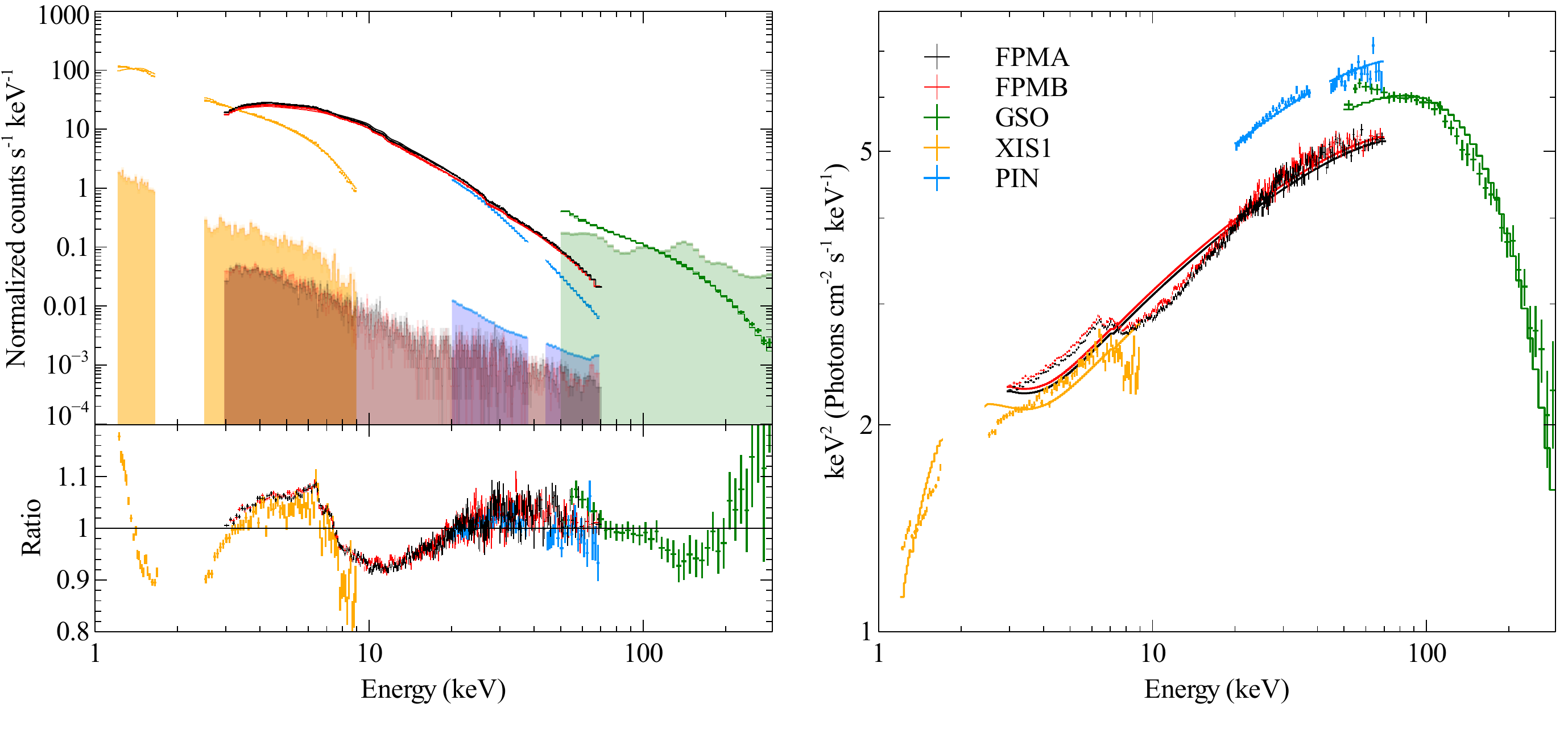}
\caption{\emph{Left}: Source and background (shaded regions) spectra from the five detectors, fit with a Comptonization model \citep[\textsc{comptt};][]{Titarchuk94}, and residuals to the fit. A constant offset is allowed between the data sets to account for flux calibration differences. \emph{Right}: The same five spectra, shown unfolded against the model, showing the sharp high energy turn over. XIS1 data are rebinned slightly in \textsc{xspec} for clarity.}
\label{fig_broadband_cutoffpl}
\end{figure*}

In the left panel of Fig.~\ref{fig_broadband_cutoffpl}, we show the five spectra fit with a Comptonization plus disk black body model (\textsc{tbabs*(diskbb+comptt)}), along with background spectra and residuals. Clear reflection features are apparent in the spectrum, along with the absorption line at $\sim6.7$~keV previously seen by \citet{Tomsick14} from the companion stellar wind. In the right panel, we show the unfolded spectrum to this model, where it becomes immediately apparent that there is a steep turnover at high energies, just above the \nustar\ energy band. This turnover has previously been observed with {\em RXTE} and \suzaku\ \citep[e.g.][]{Gilfanov00, Wilms06,Nowak11}, and appears to be similar between observations. The presence of this high energy curvature, while not immediately apparent in the \nustar\ data alone, must be taken into account before the reflection spectrum can be accurately modelled (\S~\ref{section_fullspectrum}). The model parameters for this fit are give in Table~\ref{table_contfits}, along with those for simpler power law models (with and without a cut-off). We find the best fit with the \textsc{comptt} model, which we use as our initial continuum model for the detailed spectral fitting in \S~\ref{section_fullspectrum}.

\begin{table}
\centering
\caption{Fit parameters for the continuum models (see text). We include 5~per cent systematic error on all continuum only fits, so $\chi^2$ values do not indicate absolute goodness of the fits. Cross-calibration constants are given for the best-fit model.}
\label{table_contfits}

\begin{tabular}{ l c l}
\hline
\hline
Parameter	&	Value	&	Unit/description	\\
\hline
\multicolumn{3}{l}{\textsc{Powerlaw}}\\
 	$\Gamma$	&	$1.702\pm0.003$	&	Photon index	\\
			$T_\textrm{disk}$ & $0.43\pm0.01$ & disk temperature (keV)\\
			$\chi^2/$d.o.f.	&	1973/948	&		\\
\hline
\multicolumn{3}{l}{\textsc{Cutoffpl}}\\
  $\Gamma$	&	$1.568\pm0.005$	&	Photon index	\\
			$E_\textrm{cut}$	&	$156\pm3$& High energy cutoff (keV)\\
			$T_\textrm{disk}$ & $0.53\pm0.01$ & disk temperature (keV)\\
		$\chi^2/$d.o.f.	&	941/947	&		\\
\hline
\multicolumn{3}{l}{\textsc{Comptt}}\\
  $T_\textrm{plasma}$ & $43.4\pm0.8$ & Plasma temperature (keV)\\
		 	$T_\textrm{disk}$ & $0.448\pm0.007$ & disk temperature (keV)\\
		 	$\tau$ & $1.27\pm0.03$ & Optical depth\\
		$\chi^2/$d.o.f.	&	863/947	&		\\
\hline
\multicolumn{3}{l}{Cross-normalization constants}\\
$C_\textrm{FPMA}$ & 1\\
$C_\textrm{FPMB}$ & $1.015\pm0.004$\\
$C_\textrm{XIS1}$ & $0.948\pm0.006$\\
$C_\textrm{PIN}$  & $1.31\pm0.01$\\
$C_\textrm{GSO}$  & $1.16\pm0.02$\\

\hline
\hline	
\end{tabular}
\end{table}

We also fit the more complex \textsc{eqpair} Comtponization model to these data, however with the fitting procedure outlined above this results in unphysical best-fit parameters, as the model is complex enough to be partly able to fit the reflection curvature as well as the continuum. We therefore defer discussion of this model to the more detailed fitting in \S~\ref{section_fullspectrum}.

The accuracy and reliability of the cut off measurement depends on the accuracy of the GSO background. From $\sim100$~keV upwards the GSO spectrum is background dominated, so small differences between the modelled and true background can potentially have a large effect. The background spectrum itself is completely dominated by the non X-ray background (NXB), generated by radioactive isotopes, cosmic rays, and albedo neutrons \citep{Kokubun07}. This background signal is significantly time variable, however it is well understood and can be accurately modelled and compared with the observed spectrum during Earth occultations. The estimated systematic error on the background spectrum is less than 1\% for observations of over 10~ks \citep{Fukazawa09}, which we exceed by a factor of 10. To test the systematic error this uncertainty introduces, we assume a conservative 1\% error in the normalization of the GSO background, calculate background spectra scaled up or down by this amount, and then re-fit the data using these background spectra instead. This changes the cut off energy by $\pm1.5$~keV (comparable to the $\pm0.8$~keV statistical error), but makes no qualitative difference to the spectral shape.

\subsection{Iron line fitting}
\label{section_ironline}

To avoid possible confusing effects from other spectral features we fit the XIS and \nustar\ spectra over the iron line band (4--10~keV) alone. For this section we use the full XIS exposure, rather than just the strictly simultaneous data, as changes in the continuum shape are less critical when fitting a limited energy range, and we want to maximize the signal around the iron line. There is a slight difference in the photon index measured by \suzaku\ and \nustar\ when using the full dataset, which is due to spectral variability (the simultaneous \suzaku\ and \nustar\ data agree well, and the two \suzaku\ spectra from simultaneous and non-simultaneous intervals show the same difference). We therefore allow the index to change between the two, though this is a small effect ($\sim2$ per cent), and will not affect our results or conclusions.

\begin{figure}
\centering
\includegraphics[width=\linewidth]{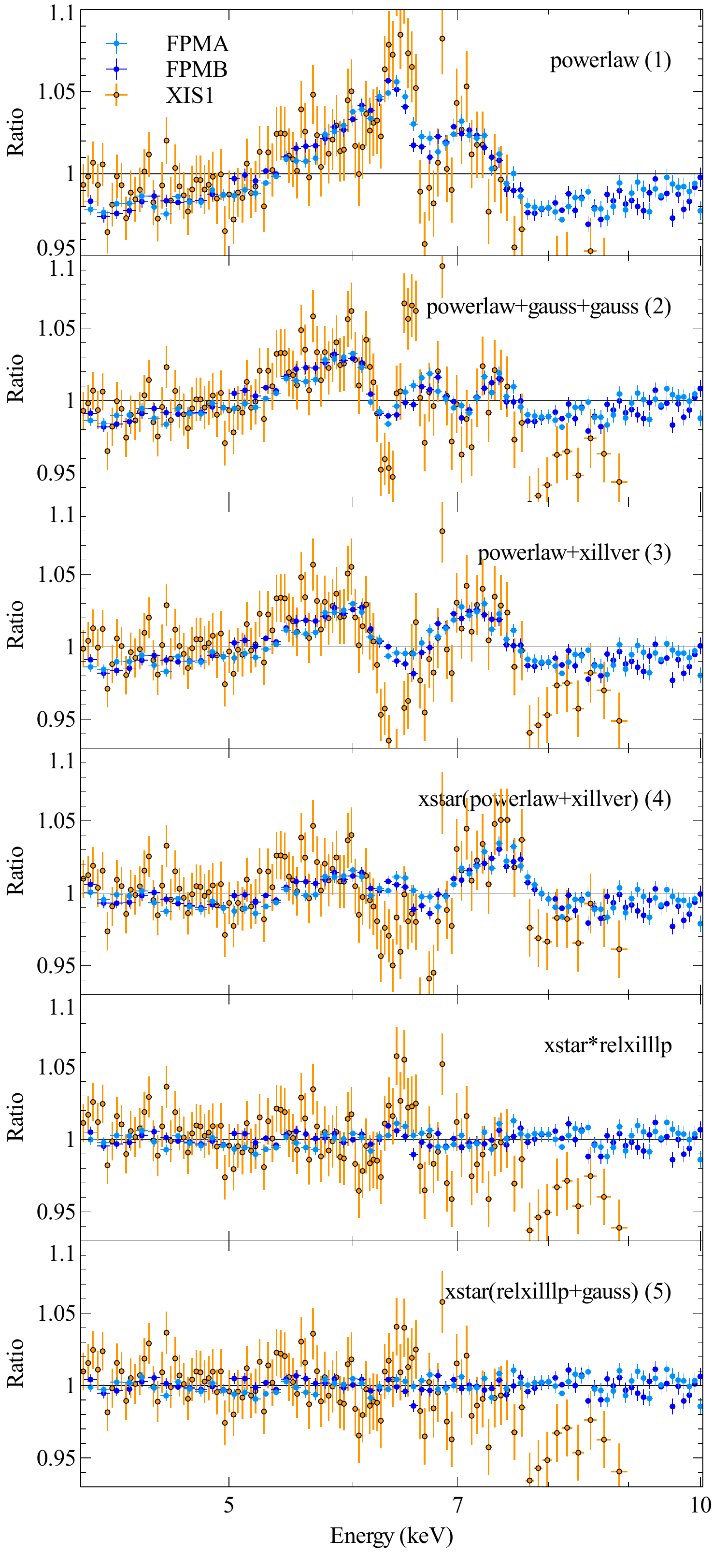}
\caption{Data to model ratios for the FPMA, FPMB 4--10~keV spectra and XIS1 4--9~keV spectrum. In all cases we include interstellar absorption. Numbers in brackets correspond to model numbers in Table~\ref{table_linefits}}
\label{fig_lineprofiles}
\end{figure}

In Fig.~\ref{fig_lineprofiles} we show the iron line profile from 4--10~keV, fit with several different models. In the top panel we show the total profile, fit with a powerlaw (including interstellar absorption). The line profile is clearly broad, extending from $\sim5$--$8$~keV, with an absorption feature at around 6.7~keV. There is a generally very good agreement between the XIS and FPM spectra, although there is a difference in flux between the two above $\sim8$~keV, which holds whether we use the full XIS spectrum or just the simultaneous data, implying it is due to calibration differences. There are known calibration issues with XIS1 spectra above $\sim8$~keV, stemming from variable background contamination from cosmic-ray photons \citep{Ishida11}, making the background level uncertain. As \nustar\ is likely to have better calibration at high energies and dominates the signal in the combined spectrum, this disagreement should not significantly affect our results.
We next fit a series of models to the line profile, shown in the remaining panels of Fig.~\ref{fig_lineprofiles}, and the model parameters are shown in Table~\ref{table_linefits}. Before fitting a blurred reflection spectrum, we test simpler models to check whether the line profile can be explained as a combination of narrow features and absorption, as the Cyg~X-1 binary system is known to have a strong stellar wind which can produce these effects \citep{Marshall01,Schulz02,Miller05}. We first fit a simple model consisting of a powerlaw plus two narrow emission lines (\textsc{powerlaw+gauss+gauss}), but this model leaves a strong, broad excess from 5--6~keV. Modelling the line with ionized, but not blurred, reflection \citep[using \textsc{Xillver}\footnote{We note that \textsc{Xillver} version 0.2d contains a bug affecting models with cut-off energies $>300$~keV, which affects both \textsc{Xillver} and \textsc{Relxill} (Garcia \& Dauser, private communication). This is not a problem for this work, as all of our models have cut-off energies $\leq 300$~keV.},][]{Garcia13} fares no better, even when we include absorption from the wind using the same \textsc{xstar} \citep{Kallman01} grid as \citet{Tomsick14}.

\begin{table}
\centering
\caption{Fit parameters for the model fits to the iron line band (4--10~keV). Model 1 is a pure power law model, model 2 includes two narrow Gaussian lines, model 3 replaces the lines with the \textsc{xillver} reflection model, model 4 includes an ionized absorber, modelled with an \textsc{xstar} grid, and model 5 is the full relativistic reflection model (see text for description). All models include interstellar absorption. Errors are only quoted when $\chi^2_\nu<2$. We give the cross-calibration constants for the best-fit model}
\label{table_linefits}
\begin{tabular}{l l c l}
\hline
\hline
Model	&	Parameter	&	Value	&	Units	\\
\hline
1	&	$\Gamma$	&	1.85	&		\\
	&	$\chi^2/$d.o.f.	&	4311/256	&		\\
\hline
2	&	$\Gamma$	&	1.85	&		\\
	&	$E_1$	&	6.34	&	keV	\\
	&   $W_1$   &   33		&	eV\\
	&	$E_2$	&	7.06	&	keV	\\
	&	$W_2$	&	20		&	eV  \\
	&	$\chi^2/$d.o.f.	&	1574/251	&		\\
\hline
3	&	$\Gamma$	&	1.87	&		\\
	&	$A_\textrm{Fe}$	&	1.01	&		\\
	&	$\log \xi_\textrm{ref}$	&	1.67	&	log(erg cm s$^{-1}$)	\\
	&	$\theta$	&	68.1	&	degrees	\\
	&	$\chi^2/$d.o.f.	&	1694/251	&		\\
\hline
4	&	$N_\textrm{H}$	&	$6.3\times10^{21}$	&	cm$^{-2}$	\\
	&	$\log \xi_\textrm{abs}$	&	3.29	&	log(erg cm s$^{-1}$)	\\
	&	$\Gamma$	&	1.88	&		\\
	&	$A_\textrm{Fe}$	&	1.01	&		\\
	&	$\log \xi_\textrm{ref}$	&	1.67	&	log(erg cm s$^{-1}$)	\\
	&	$\theta$	&	68.1	&	degrees	\\
	&	$\chi^2/$d.o.f.	&	1395/249	&		\\
\hline
5	&	$n_\textrm{H}$	&	$1.47_{-0.17}^{+0.08}\times10^{22}$	&	cm$^{-2}$	\\
	&	$\log \xi_\textrm{abs}$	&	$4.85\pm0.02$	&	log(erg cm s$^{-1}$)	\\
	&	$h$	&	$<1.21$	&	$r_\textrm{ISCO}$	\\
	&	$a$	&	$>0.84$	&		\\
	&	$r_\textrm{IN}$	&	$1.11\pm0.01$	&	$r_\textrm{ISCO}$	\\
	&	$\Gamma$	&	$1.926_{-0.006}^{+0.001}$	&		\\
	&	$A_\textrm{Fe}$	&	$<0.55$	&		\\
	&	$\log \xi_\textrm{ref} $	&	$2.768\pm0.003$	&	log(erg cm s$^{-1}$)	\\
	&	$\theta$	&	$47.8\pm0.3$	&	degrees	\\
	&	$R$	&	$0.809\pm0.01$	&		\\
	&   $E_\textrm{Gauss}$ & $6.43\pm0.01$ & keV   \\
	&	$W_\textrm{Gauss}$ & $12\pm1$ & eV\\
	&	$\chi^2/$d.o.f.	&	294/249	&		\\
\hline
\multicolumn{3}{l}{Cross-normalization constants}\\
&$C_\textrm{FPMA}$ & 1\\
&$C_\textrm{FPMB}$ & $1.0157\pm0.0007$\\
&$C_\textrm{XIS1}$ & $1.148\pm0.008$\\
\hline
\hline	
\end{tabular}
\end{table}

The exceptional quality of the combined spectrum makes it clear that only a relativistically blurred iron line can explain the observed line profile. The broad excess remaining from 5--6~keV after fitting the two Gaussian model must be due to blurred reflection, as there are no strong atomic lines in this band. We fit the relativistic reflection spectrum with the \textsc{relxilllp} model \citep{Garcia14}, which self-consistently calculates the reflected emission as a function of inclination for an on-axis point source above the disk. Relative to models which assume a broken power law emissivity profile, this has the advantage of parametrizing the profile in terms of a physical quantity, reducing the number of parameters needed to describe the profile from 3 to 1, and restricting the profile to regions of parameter space that make physical sense. In addition to the broad line, we find a weak narrow emission feature at 6.4~keV which we model with a narrow Gaussian line (see bottom 2 panels). This feature is only obvious in the full XIS spectrum, but corresponds to an excess in the \nustar\ spectrum as well, and results in a highly significant improvement in the fit statistic $\Delta\chi^2=56$ for 2 additional degrees of freedom. The iron line in Cyg~X-1 was shown to be a composite of broad and narrow lines using the \chandra\ High Energy Transmission Grating Spectrometer (HETGS) by \citet{Miller02}. The best fit model parameters are shown in Table~\ref{table_linefits}.

\begin{figure*}
\centering
\includegraphics[width=0.8\linewidth]{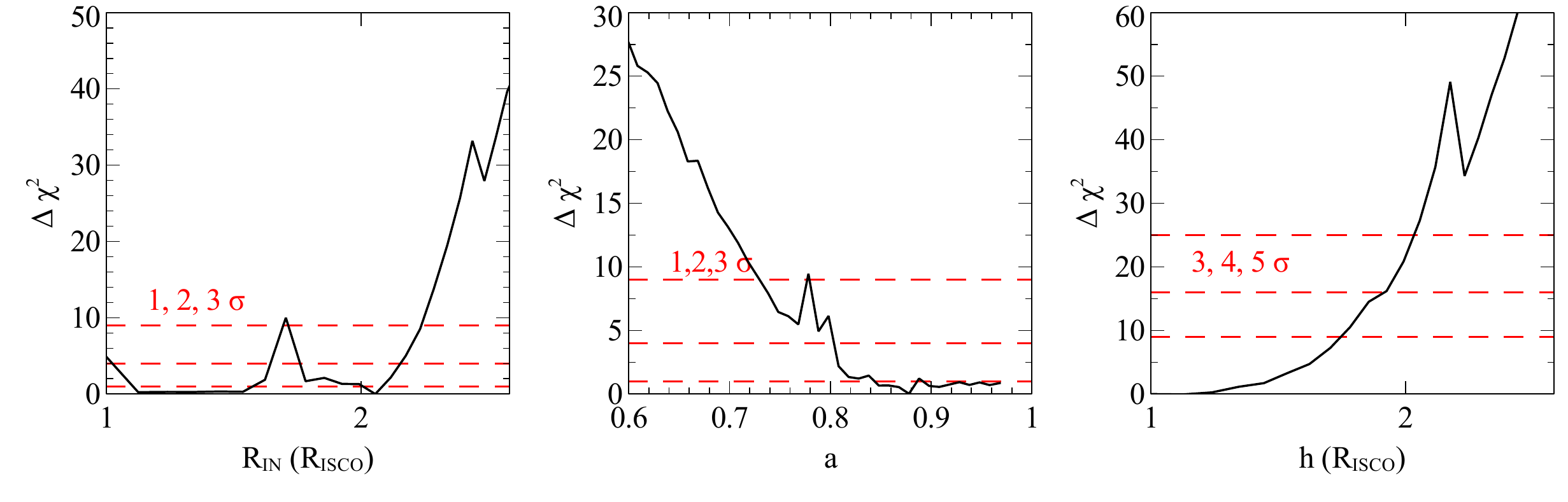}
\caption{\emph{Left}: $\chi^2$ contour plot for the inner radius of the accretion disk, in units of the ISCO. \emph{Middle}: $\chi^2$ contour plot for black hole spin parameter. \emph{Right}: contour plot for the height of the emitting source above the disk plane, in units of the ISCO radius. All plots are calculated using the \textsc{steppar} function in \textsc{Xspec}.}
\label{fig_lineprofile_rin_chisq}
\end{figure*}

In Fig.~\ref{fig_lineprofile_rin_chisq} we show $\Delta\chi^2$ contours from fitting the iron line profile for the spin ($a$), source height ($h$), and inner radius ($r_\textrm{IN}$) model parameters. All three are strongly constrained to values indicating strongly blurred emission from the innermost regions of the accretion disk. Even if we assume the 3$\sigma$ limits on $a$ and $r_\textrm{IN}$ (0.8 and 2~$r_\textrm{ISCO}$) this gives a inner radius of only $\sim6 r_\textrm{G}$, which argues strongly against significant disk truncation.

We check for degeneracy between parameters by using a Markov chain Monte Carlo (MCMC) algorithm to explore the parameter space\footnote{We use the \textsc{xspec\_ emcee} code by Jeremy Sanders, based on the \textsc{emcee} python implementation \citep{Foreman-Mackey13} of the Goodman-Weare affine invariant MCMC ensemble sampler \citep{Goodman10}.}. We use 50 walkers with 15,000 iterations each, burning the first 5,000. The walkers are initially distributed around the best fit value found by \textsc{Xspec}, following a normal distribution in each parameter, with the standard deviation set to the delta value of that parameter (we ensure that the delta values are all smaller than the uncertainties on each parameter). We check that the auto-correlation time is at least 100 times smaller than the chain length for all parameters, and `jack-knife' the chains (comparing the distributions from the first and second halves of the chain) to check that there are no large systematic differences between the distributions. 

\begin{figure*}
\centering
\includegraphics[width=\linewidth]{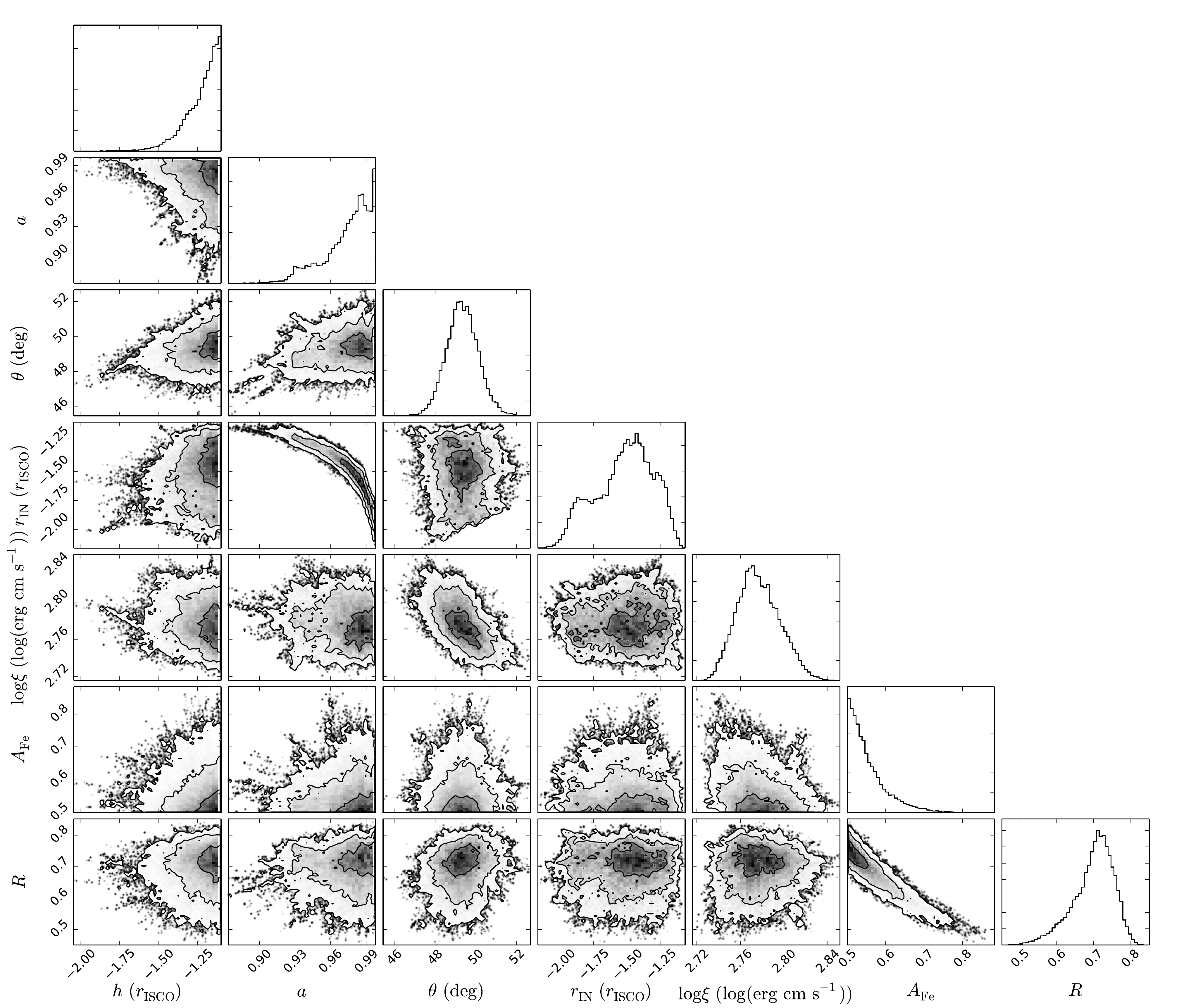}
\caption{Output distributions from the MCMC analysis of the best fit model of the iron line profile. Contours correspond to 1, 2 and 3 $\sigma$. Points are shown where the density of points drops beyond the $3\sigma$ limit. $y$-axes for the histograms are in arbitrary units. Only two of the contour plots show strong degeneracy: $a$ and $r_\textrm{in}$  are degenerate, as are the iron abundance and reflection fraction. Negative values for $r_\textrm{in}$ and $h$ correspond to units of $r_\textrm{ISCO}$, following the convention from \textsc{Relxill}.}
\label{fig_lineprofile_contours}
\end{figure*}

We calculate the distributions of all parameters (Fig.~\ref{fig_lineprofile_contours})\footnote{This plot and the plot in Fig.~\ref{fig_broadband_contours} are produced using the \textsc{Triangle.py} python module \citep{Foreman-Mackey14}}, and find significant degeneracies in only two cases. The truncation radius of the disk is degenerate with the spin, as expected as both parameters control the effective inner radius. While these parameters cannot therefore be independently constrained, this remains a very strong limit on the truncation of the inner disk. If we take the $3\sigma$ limits of the distribution, at no point does the inner radius exceed $3 r_\textrm{G}$, so this strongly requires the disk not to be truncated\footnote{This differs from the 6~$r_\textrm{G}$ previously quoted because that estimate did not take into account parameter degeneracies}. 

Additionally, the reflection fraction and iron abundance are degenerate. This is also expected since we are only fitting the iron line band, so a true constraint on either value is hard to achieve - within this limited energy range, the main effect of both of these parameters is to alter the strength of the iron line. This goes some way toward explaining the sub-solar iron abundance we find, which is unlikely to be correct as the iron abundance in Cyg-X-1 is generally thought to be super-solar \citep{Hanke09,Duro11,Fabian12,Tomsick14}. We note that there is a slight discrepancy between the \textsc{steppar} and MCMC contours - for example, the 3-$\sigma$ limit on $a$ is lower when calculated using \textsc{steppar}. This difference can be explained by the limited accuracy possible with \textsc{steppar} - while both methods agree on the requirement for high spin, the best fit found by \textsc{steppar} has a $\chi^2$ value of 298, compared to the solution with 294 found by the MCMC analysis. In addition, the power law index and ionization parameter $\xi_\textrm{ref}$ are degenerate, although this covers only a very small range in $\log \xi_\textrm{ref}$.

To investigate the iron abundance further, we re-fit the data with the abundance fixed at 2, as found by \citet{Tomsick14}. This  worsens the fit significantly ($\Delta\chi^2=15$ for one fewer degree of freedom), however it introduces only minor differences in the other parameters (all of the relativistic blurring parameters are consistent within errors). The only parameter to change significantly is the reflection fraction, which decreases from $0.81\pm0.01$ to $0.34\pm0.01$. Because of the lack of change in other parameters and absence of significant residual features it seems likely that the constraint on the iron abundance is being driven by random statistical fluctuations in the data or a minor feature of the reflection model. 

\begin{figure}
\centering
\includegraphics[width=\linewidth]{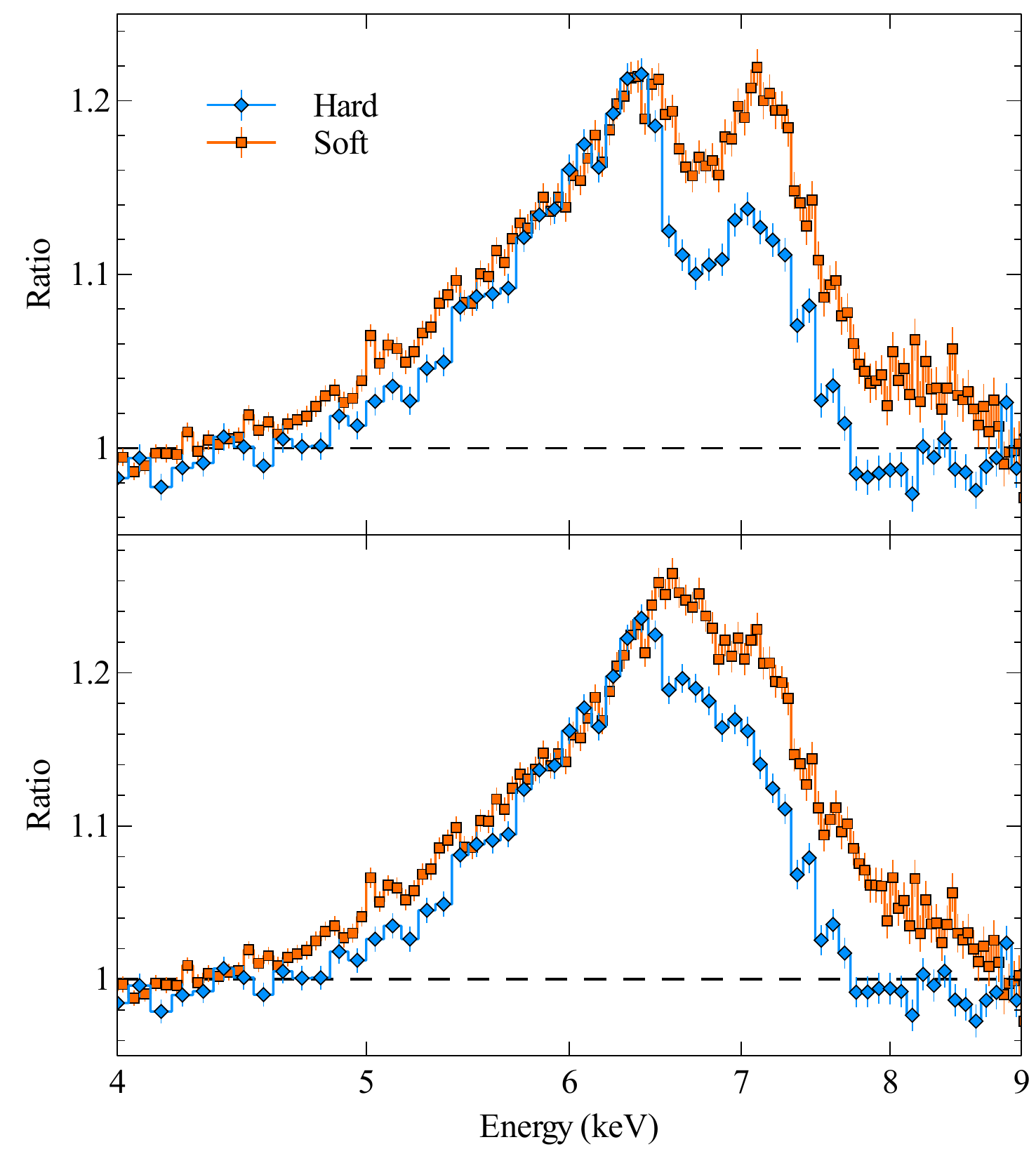}
\caption{\emph{Top}: Comparison of the \nustar\ iron line profiles of Cyg~X-1 in the hard and soft states, with observations taken at approximately the same orbital phase. The hard state profile is scaled to approximately match the flux of the soft state line at 6~keV. \emph{Bottom}: As top, but after correcting for the $\sim6.7$~keV absorption line using the parameters from the best fit models in \citet{Tomsick14} and this work.}
\label{fig_lineprofile_comparison}
\end{figure}

In the top panel of Fig.~\ref{fig_lineprofile_comparison} we show a comparison of the iron line profiles from the soft and hard states. We use the soft state observations 30001011002 and 30001011003, which were presented by \citet{Tomsick14}, and were taken at a similar orbital phase (0.845--0.962 in \citeauthor{Tomsick14} compared to 0.902--0.104 in this work, so the absorption should be similar). We plot the combined FPMA and FPMB spectrum from both observations. These data are reduced and extracted as in \citet{Tomsick14}. We fit both spectra with simple continuum models over the 3--4 and 8--10~keV bands. For the hard state, we use a power law, and in the soft state a power law plus disk black body, with the temperature fixed at the best fit value from \citet{Tomsick14}. We then scaled the residuals of the hard state by a factor of 3 so that the amplitudes of the two lines are comparable. It is immediately apparent from the residuals that there are no significant qualitative changes to the red wing of the line, which is the part of the line profile which should be most affected by truncation of the disk. In both cases, the line extends to below $\sim5$~keV. The $\sim6.7$~keV absorption feature appears to be consistent between the two, with the spectral changes concentrated on the blue side of the line. In the  lower panel we show the same profiles, but after correcting for the absorption from the companion wind (using the absorption parameters from model 5). This makes the 6.4~keV excess from the narrow line in the hard state more obvious, and demonstrates more clearly the lower energy of the peak in the hard state. 

\subsection{Full spectrum fits}
\label{section_fullspectrum}
We now fit all five spectra (XIS, FPMA, FPMB, PIN and GSO) simultaneously over the full energy range, modelling both the continuum and reflection components. We allow for a constant normalization offset to account for flux calibration differences between the spectra. As in \S~\ref{section_continuum} we use only the XIS1 data that was taken simultaneously with \nustar .

We initially fit the full spectrum with a combination of the models for the continuum (\textsc{comptt + diskbb}; \S~\ref{section_continuum}) and line profile (\textsc{Relxill + gauss}; \S~\ref{section_ironline}). The full model is, including absorption, \textsc{tbabs * xstar * (diskbb + comptt + relxillp + gauss)}. As the \textsc{relxill} models assume a cut-off power law as the illuminating spectrum, we fix the cut-off of the reflection spectrum to three times the plasma temperature of the Comptonization component (i.e. the peak of the Wien law in energy flux), which is approximately the cut-off energy that would be measured with a power law model. This model does not result in a formally acceptable fit - we find no solutions with $\chi^2_\nu<2$. It is not clear why this model does not work, as it results in significant residuals over the full energy band. We conclude that it is likely due to subtle spectral curvature that cannot be accounted for by the simple continuum model used. We test adding a partial-covering ionized absorption zone as a potential explanation for the curvature, and find that while it significantly improves the fit ($\Delta\chi^2\sim500$, for 3 additional degrees of freedom) the fit is still not acceptable ($\chi^2_\nu=1669/839=1.99$)

\begin{figure*}
\centering
\includegraphics[width=\textwidth]{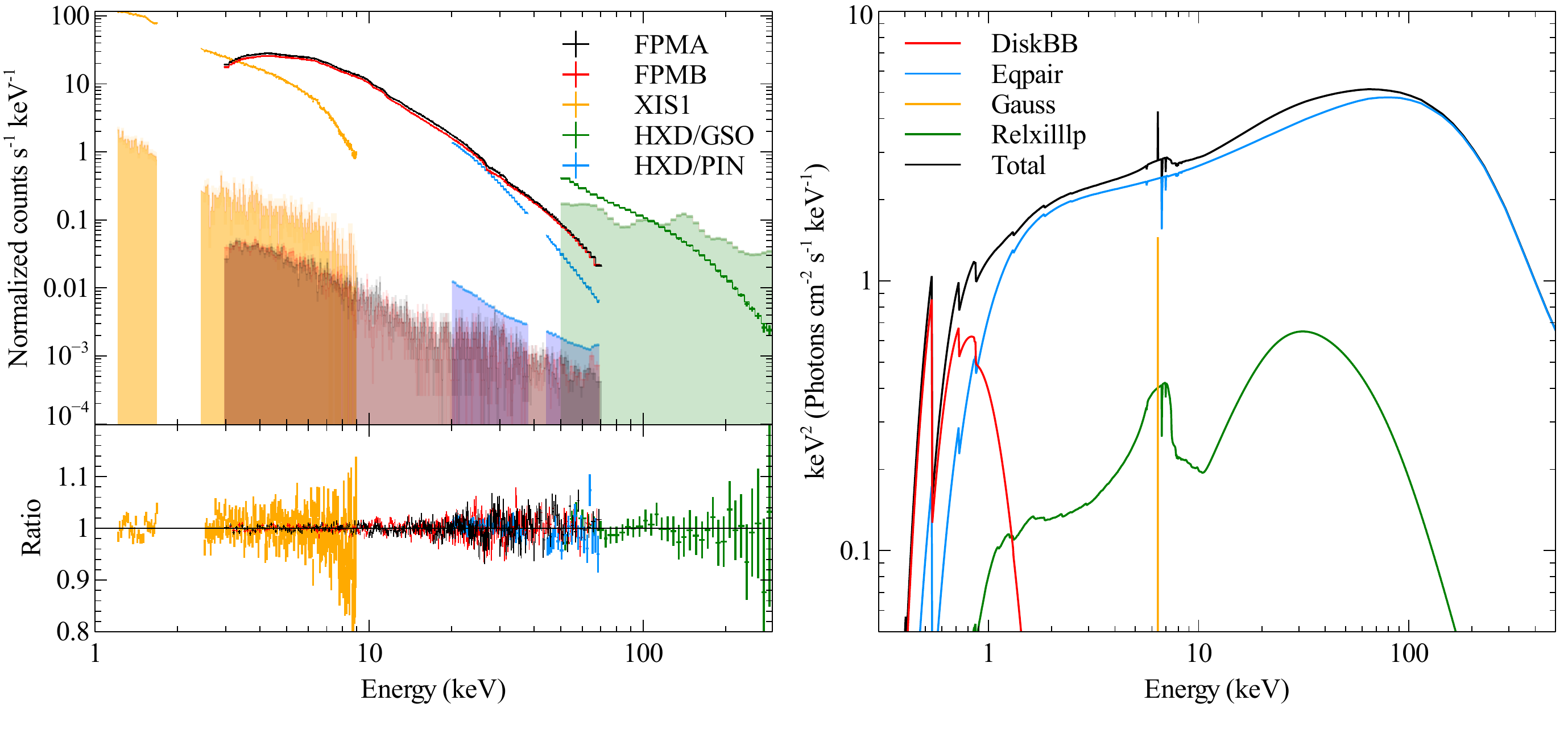}
\caption{\emph{Left}: Upper panel shows the broad-band spectrum and best fit \textsc{eqpair} model; lower panel shows the ratio of the data to the model. Shaded regions show the background spectra for each instrument. \emph{Right}: The best fit model, showing the different model components.}
\label{fig_broadband_bestfit}
\end{figure*}

Next, we fit the more sophisticated \textsc{eqpair} model \citep{Coppi99}. \textsc{eqpair} calculates the Comptonized spectrum for a combined population of thermal and non-thermal electrons. \textsc{eqpair} was used extensively to model the broad band spectrum of Cyg~X-1 by \citet{Nowak11}, who found that it could account for the spectral shape well. As in \citeauthor{Nowak11} we assume the default distribution of non-thermal electrons: a density following a power law $\gamma^{-2}$ between $\gamma_\textrm{min}=1.3$ and $\gamma_\textrm{max}=1000$, where $\gamma$ is the Lorentz factor. While this model can also calculate the reflection spectrum as of version 1.10, we set the flux of this component to zero and continue to use \textsc{relxilllp} for our reflection model. This is done for two reasons: firstly to maintain consistency with the other models and the fits in \S~\ref{section_ironline}, and secondly because the \textsc{relxill} models are more sophisticated and include more atomic physics than the \textsc{ireflct} model used by \textsc{eqpair}. We fix the cut-off of the illuminating spectrum of the reflection model to 100 keV, as there is no parameter to tie it to in the \textsc{eqpair} model, and it is very hard to constrain independently. We choose 100~keV as it corresponds to the peak of the spectrum shown in Fig.~\ref{fig_broadband_cutoffpl}, and is in the range of 2--3 times the plasma temperature (Table~\ref{table_contfits}) expected for a powerlaw approximation to a Comptonization spectrum. While this is not an exact value, the systematic error introduced is unlikely to be large, comparable to that introduced by the reflection model assuming a power law as the input spectrum.
Following \citet{Nowak11}, we include an un-Comptonized black body as well as the Comptonized one included in the \textsc{eqpair} model. This model can better account for the subtle spectral curvature, and results in a much better fit to the broad band spectrum ($\chi^2=1219/919=1.33$). The data, model and residuals are shown in Fig.~\ref{fig_broadband_bestfit}, and the parameters are given in Table~\ref{table_broadband}.

\begin{table}
\caption{Fit parameters and normalization constants for the best fit broad band model. $l_\textrm{h}$,$l_\textrm{s}$, and $l_\textrm{nt}$ refer to the hard, soft, and non-thermal compactnesses, respectively. The X-ray luminosity is absorption corrected, however the fluxes are not. We assume the distance of 1.86~kpc from \citet{Reid11} and mass of 14.8 $M_\odot$ from \citet{Orosz11}. The reflection fraction $R$ is calculated from the ratio of model fluxes, rather than the model parameter, hence no error is calculated.}
\label{table_broadband}

\begin{tabular}{l c l}
\hline
\hline
Parameter & Value & Units\\
\hline
\multicolumn{3}{l}{Interstellar absorption (\textsc{tbabs})}\\
$N_\textrm{H}$ & $(5.1\pm0.1)\times10^{21}$   & cm$^{-2}$\\
\hline
\multicolumn{3}{l}{Wind absorption (\textsc{xstar})}\\
$N_\textrm{H}$         & $(1.35\pm0.07)\times10^{22}$ & cm$^{-2}$\\
$\log(\xi_\textrm{abs})$    & $4.81\pm0.02$				 & log(erg cm s$^{-1}$)\\
\hline
\multicolumn{3}{l}{disk black body (\textsc{diskbb})}\\
$T_\textrm{BB}$			  & $0.104^{+0.016}_{-0.004}$	 	 & keV\\
\hline
\multicolumn{3}{l}{Comptonization (\textsc{eqpair})}\\
$l_\textrm{h}/l_\textrm{s}$  & $0.75^{+0.06}_{-0.08}$\\
$l_\textrm{nt}/l_\textrm{h}$ & $0.060_{-0.010}^{+0.003}$\\
$\tau$						  & $0.46\pm0.01$\\
\hline
\multicolumn{3}{l}{Narrow line (\textsc{gauss})}\\
$E_\textrm{Gauss}$   		  & $6.40\pm0.01$				 & keV\\
$W_\textrm{Gauss}$ 			& $10\pm1$ 				& eV\\
\hline
\multicolumn{3}{l}{Reflection (\textsc{relxilllp})}\\
$h$						  & $<1.56$						 & $r_\textrm{ISCO}$\\
$a$ 						  & $>0.97$\\
$r_\textrm{in}$			  & $1.5\pm0.3$			 	 & $r_\textrm{ISCO}$\\
$\theta$					  & $45.3\pm0.4$				 & degrees\\
$\Gamma$					  & $1.60\pm0.02$				 &\\  
$\log(\xi_\textrm{ref})$	  & $3.03\pm0.01$				 & log(erg cm s$^{-1}$)\\
$A_\textrm{Fe}$			  & $4.7\pm0.1$\\
$R$						  & 0.16\\
\hline
\multicolumn{3}{l}{Cross-normalization constants}\\
$C_\textrm{FPMA}$ & 1\\
$C_\textrm{FPMB}$ & $1.0163\pm0.0005$\\
$C_\textrm{XIS1}$ & $0.923\pm0.002$\\
$C_\textrm{PIN}$  & $1.295\pm0.003$\\
$C_\textrm{GSO}$  & $1.192\pm0.005$\\
\hline
\multicolumn{3}{l}{Fluxes}\\
2--10~keV (XIS1) & $7.57\times 10^{-9} $ & erg~cm$^{-2}$~s$^{-1}$\\
2--10~keV (FPMA) & $6.35\times 10^{-9} $ & erg~cm$^{-2}$~s$^{-1}$\\
10--50~keV (FPMA) & $1.04\times 10^{-8} $ & erg~cm$^{-2}$~s$^{-1}$\\
$L_{\textrm{X,0.1--200}}/L_\textrm{Edd}$ (FPMA)& 0.0250\\
\hline
\hline
\end{tabular}
\end{table}

We note that in the best fit model almost all the black body flux is outside the range covered by the regions of the XIS1 spectrum that we fit. It is therefore likely that the temperature is being affected more strongly by the \textsc{eqpair} continuum component than the black body itself. The measured temperature, while consistent with the hard state in Cyg~X-1 \citep{Balucinska95,DiSalvo01}, should therefore be treated with a degree of caution.

\begin{figure*}
\centering
\includegraphics[width=\linewidth]{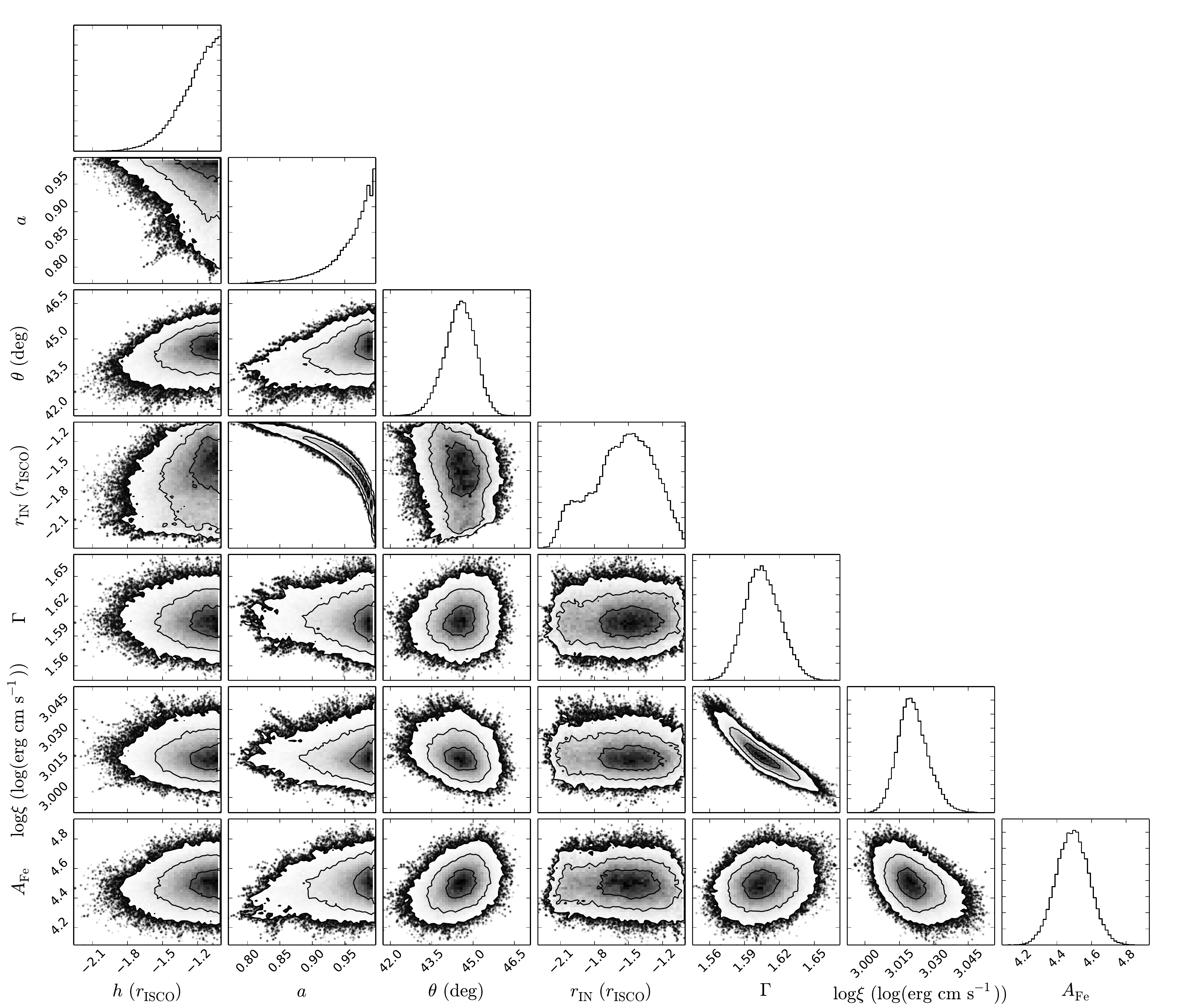}
\caption{Output distributions from the MCMC analysis of the best fit model of the broad band spectrum of Cyg~X-1. Contours correspond to 1, 2 and 3 $\sigma$. Points are shown where the density of points drops beyond the $3\sigma$ limit. $y$-axes for the histograms are in arbitrary units. Negative values for $r_\textrm{in}$ and $h$ correspond to units of $r_\textrm{ISCO}$, following the convention from \textsc{Relxill}.}
\label{fig_broadband_contours}
\end{figure*}

We show the confidence contours for the reflection spectrum in Fig.~\ref{fig_broadband_contours}. As in \S\ref{section_ironline}, these are calculated using MCMC chains. As we are exploring a larger parameter space, we increase the number of walkers to 100, and the chain length to 20,000. The results are very similar to those found from fitting the line profile alone - the blurring parameters are strongly constrained to produce a high level of blurring, with a small inner radius, low source height, and high spin. Again, the spin and inner radius are highly degenerate, but are inconsistent with a large degree of disk truncation, with $r_\textrm{in} < 3r_\textrm{G}$ in all cases.

The \textsc{eqpair} continuum model includes a Comptonization of a population of non-thermal electrons. We test whether a hybrid electron population is important to the fit by fixing the compactness of non thermal electrons, $l_\textrm{nt}$, to zero. We find a significantly worse fit with this model ($\Delta\chi^2=222$, for one additional degree of freedom), which indicates that non-thermal scattering is important in the continuum modelling.

\section{Discussion}
\label{section_discussion}

As found by \citet{Tomsick14}, the inclination values we measure based on the reflection spectrum are not consistent with the binary inclination of $27.1\pm0.8$ degrees from \citet{Orosz11}, based on optical measurements. \citeauthor{Tomsick14} use a selection of models and find inclinations in the range 42--69 degrees. Progressing from the recent work of \citeauthor{Tomsick14}, we are using the latest reflection models which self-consistently calculate the reflection spectrum and relativistic blurring as a function of angle \citep{Garcia14}, but we nevertheless find a similarly high inclination of $45.3\pm0.4$ degrees from our best fit broadband model. \citeauthor{Tomsick14} discuss the high ionization value found in the soft state as a possible cause of higher inclinations, due to additional Compton broadening of the line not accounted for by the model. This is unlikely to be the same in the hard state, which we find has $\sim20$ times lower ionization than the soft state, so we would expect significantly less broadening and a correspondingly lower inclination value. Including the expected broadening
\begin{equation}
\frac{\sigma}{E} \simeq \left( \frac{2kT_\textrm{BB}}{m_\textrm{e}c^2} \right)^{1/2}
\end{equation}
for a given disk temperature $T_\textrm{BB}$ and energy $E$ \citep{Pozdnyakov83}, does not significantly affect either the quality of the fit, or the inclination.
As a high inclination value has now been found with \nustar\ in two observations with different reflection models and in different states this rules out a range of systematic effects, making it more likely that there is a real physical effect behind this measurement. The receding jet in Cyg~X-1 has not been detected \citep{Stirling01}, which points to a small inclination angle, however the jet angle, which would offer an independent constraint, is not strongly constrained. It is possible that the differences between the binary inclination and our inclination are due to precession of the disk or a warp in the disk. A strong modulation with a period of $142\pm7$ days is identified as the effect of precession by \citet{Brocksopp99}, with an angle of up to 37 degrees between the binary plane and the disk. This would be easily enough to explain the observed discrepancy, however there are uncertainties in the precession measurement \citep{Zdiarski11}, and if this is the case we would expect the measured inclination to change with this period, which has not so far been observed. Additionally, this period has only been observed in the hard state so may not be able to explain the soft state result. A potential problem with a warped disk is that  thick disks are not expected to align \citep{Ivanov97,Fragile09}, so this may not be possible, depending on the state of the accretion flow.

We find some minor differences in the results when we fit the Fe K$\alpha$ line alone compared to when we fit the full spectrum. The $\Gamma$ value found for the reflection component in the broad band model, where it is independent of the continuum (\S~\ref{section_fullspectrum}), is significantly lower than that found when fitting a power law to the low energy continuum (\S~\ref{section_ironline}), although this does not appear to impact the other reflection parameters. If, as has been suggested elsewhere \citep{Fabian14}, the continuum spectrum seen by the disk differs from that seen by an external observer, then tying the $\Gamma$ values of the reflection and continuum components may not be appropriate (this has also been observed in GX~339-4 by F\"urst et al., submitted). Additionally, we find a large difference in the iron abundance between the two fits, with the line fit returning a very low abundance ($<0.51$ solar) and the broad band fit giving a super-solar value ($4.7\pm0.1$). The estimate from the line profile fit cannot be treated as reliable, as a measurement of the iron abundance requires additional reflection features to establish the relative strength of the iron line. The result from the broad-band fit is more plausible, but could still be affected by systematic effects such as the assumed density of the disk\footnote{Changes in the density can potentially alter the strength of the iron line relative to the soft excess and Compton hump without otherwise changing the shape of the line profile. A discussion of this and other density effects will be presented in future work (Fabian et al., in prep.)}. We note additionally that neither of the iron abundances found in this work is consistent with the value from \citet{Tomsick14}. This difference cannot be due to changes in the iron abundance of Cyg~X-1, and must therefore be due to systematic effects unaccounted for by the modelling.

The strength of the relativistic blurring in the hard state spectrum argues strongly against the disk being truncated. The line profile is not significantly narrower in the hard state compared to the soft state, and our best fitting models exclude an inner disk radius larger than $3r_\textrm{G}$ at $3\sigma$, which does not allow for significant truncation in the switch from soft to hard states. This holds whether we fit the whole spectrum, or restrict our fits to the line profile. A similarly constant line profile was found by \citet{Reis11} in XTE~J1752-223, where the shape of the line profile was not significantly affected by the state transition. The main difference between the hard and soft line profiles appears to be attributable to changes in the ionization parameter of the underlying reflection spectrum, rather than the extent of the relativistic blurring. The blurring parameters we find in this work are generally consistent with those found by \citet{Tomsick14}, but the ionization parameter differs by an order of magnitude and can strongly affect the shape of the line profile, extending it to higher energies \citep[e.g.][]{Ross05}.
If the disk is not truncated, another explanation for the lower reflection fraction in the hard state is needed: the ratio of reflected to continuum fluxes in the 20--40~keV band is 0.16 from our best fitting model, and 0.52 for model 7 from \citet{Tomsick14}. Potential explanations include outflowing material \citep{Beloborodov99,Miller12} or increased Comptonization of the reflected emission by the corona \citep{Wilkins15}.
We find some changes in the shape of the iron line profile between the soft and hard states, although we do not make a detailed comparison. These differences are largely on the higher energy side of the line - the red wings of the two line profiles overlap very closely when the hard state residuals are scaled to match those of the soft state. A more detailed investigation of the changes between the states will be presented in in future work.

Measuring the spin or inner radius of XRBs in the hard state is significantly harder than in the soft state due to the reduced flux and lower reflection fraction. For this reason there are relatively few estimates of these parameters to compare our results to, and no strong consensus on true values. Our result ($a>0.97$) is consistent with recent measurements of the spin in the hard state by \citet{Miller12} ($0.6<a<0.99$) and \citet{Fabian12} ($0.97^{+0.014}_{-0.02}$). Previous works on the hard state have been unable to simultaneously measure the spin and inner radius, instead using one as a proxy for the other, due to the strong degeneracies and limited spectral quality available. Our new measurement thus represents a significant step forward in understanding the behaviour of Cyg~X-1 in the hard state, made possible by high quality \nustar\ spectra and the latest generation of reflection models. These spin values are also consistent with the near maximal values found in the soft state, where the line profile is clearer and the spin can be independently measured using the continuum fitting method \citep{Gou11,Gou14,Tomsick14}.

We use the \textsc{relxilllp} relativistic reflection model throughout this work. Unlike other such models, \textsc{relxilllp} explicitly assumes a particular geometry - that of an on-axis point source. In reality, the corona with have some radial and vertical extent, so eventually this assumption will break down. However, this should not affect the results presented here, as the strength of the relativistic blurring is very high. The combination of high spin, low source height and small inner disk radius all point toward a compact source, close to the event horizon. Under these circumstances the point source assumption is valid beyond the limits of current instrumentation. Introducing either a large radial or vertical extent to the source would flatten the emissivity profile \citep{Wilkins12,Dauser13}, reduce the level of blurring and worsening the fit. One possibility for an alternative geometry that retains the high level of blurring would be to introduce an outflowing vertically extended corona, where the continuum emission comes from large height regions and the reflection spectrum is produced by the emission from the base of the corona. However this is a complex geometry to model, making it hard to implement and beyond the scope of this work. In any case, the requirements for high spin and small inner disk radius would have to be retained in any such model, otherwise the broad line could not be produced.

The weak narrow line detected in the \nustar\ and XIS1 spectra has been detected previously in the hard state with \chandra\ \citep{Miller02}, but was not detected in the soft state by either \chandra\ \citep{Feng03} or in the combined \nustar\ and \suzaku\ soft state observation \citep{Tomsick14}. This difference is most likely due to the lower flux level in the hard state and the smaller contribution from relativistic reflection, which is typically much stronger in the soft state. The equivalent width of the line we find, $10\pm1$~eV, is lower than that found by \citet{Miller02}, $16^{+3}_{-2}$~eV, but not by a great deal. This could indicate changes in the outer disk or stellar wind, where some of the line flux is thought to originate, or could be due to systematic effects introduced by the different levels of absorption in the observations - the \chandra\ spectrum was taken when there was very little absorption by the stellar wind.

Simple phenomenological cut-off power law models for the continuum spectrum of Cyg~X-1 fail to properly reproduce the sharp high energy cut-off observed. This is a known problem with these models \citep[e.g.][]{Zdziarski03}, as the turnover predicted by Comtonization can be significantly steeped than an exponential. We find that a simple disk black body, thermal Comptonization and reflection model cannot adequately describe the broad-band spectrum of the source. There are several potential explanations for this, which cannot be easily differentiated. The Comptonization model used may not be suitable for fitting the spectrum \citep[\textsc{comptt} is only an analytical approximation of thermal Comptonization;][]{Titarchuk94}. We find a significantly improved fit using the more complex \textsc{eqpair} model, which includes both thermal and non-thermal Comptonization. Alternatively, the additional curvature may be due to deficiencies in the reflection modelling. The \textsc{Relxilllp} model used here is calculated for a density of $10^{15}$~cm$^{-3}$, which is closer to the expected density of an AGN accretion disk than that of an XRB. Differences in the density of the reflecting medium are unlikely to make a difference to the parameters returned for the relativistic blurring of the iron line, but can cause subtle changes to the broad-band reflection spectrum (Fabian et al., in prep). 

We find that the same best-fit \textsc{eqpair} model gives a significantly worse fit to the data when we exclude non-thermal electrons. This, combined with the failure of the \textsc{comptt} model, demonstrates that thermal Comptonization alone cannot describe the continuum of Cyg~X-1 in the hard state. This leaves two possibilities - a combination of thermal and non-thermal Comptonization as in our best fit-model, or a jet based model \citep{Markoff05,Maitra09}. Although we do not test jet descriptions here, we demonstrate that the high quality broad band observations of XRBs now possible with \nustar\ can distinguish between different continuum models. This is a promising avenue for further research. If the Comptonization interpretation is correct, this may provide a possible explanation for the difference in the measured continuum spectrum and that inferred from the reflection spectrum. The reflected emission could potentially be produced primarily from one population only, leading to the observed difference.

\section{Conclusions}
\label{section_conclusions}
The \nustar\ and \suzaku\ spectra of the hard state in Cyg~X-1 have resulted in a broad band spectrum of exceptional quality, which enables weak reflection features to be measured more precisely than ever before. Our main conclusions are summarised below:

\begin{itemize}
\item The broad-band spectrum of Cyg~X-1 shows a very steep cut-off above $\sim100$~keV, detected with the GSO, corresponding to a plasma temperature of $43.4\pm0.8$~keV. This cut-off is not apparent in the \nustar\ spectrum alone and is worth considering in other \nustar\ spectra of hard-state binaries, where simultaneous GSO data may not be available to measure the cut-off energy.
\item Fitting the iron line energy band alone (from 4--10~keV) we find that the line cannot be modelled with a combination of narrow lines and ionized absorption, and that a relativistic line profile is needed. In addition, we find a weak narrow line is needed to fully describe the profile.
\item When we carefully model the relativistic blurring of the iron line, we find that it requires a high degree of smearing, with a high spin ($a>0.97$) and small inner radius ($r_\textrm{in}<1.7$~$r_\textrm{ISCO}$), as well as a low source height ($h<1.56$~$r_\textrm{ISCO}$). This rules out significant truncation of the disk at the transition to the hard state. While the disk remains close to the ISCO in the bright part of the hard state, disk truncation at much lower Eddington fractions remains possible.
\item A simple comparison of the line profiles from the soft and hard states shows that the line has not become significantly narrower. While the high energy side of the line appears to drop earlier in the hard state, the red wing extends to similarly low energies, and the differences can be attributed to differences in the ionization of the disk.
\item We confirm the high level of blurring with fits to the full spectrum, simultaneously modelling the continuum and reflection spectra. After running extensive MCMC chains to explore the parameter space, we do not find any solutions with $r_\textrm{in}>3r_\textrm{G}$ at the $3\sigma$ confidence level.
\item We find that thermal Comptonization models (\textsc{comptt} and \textsc{eqpair} without non-thermal electrons) cannot adequately describe the continuum. Instead a hybrid model is required for Comptonization to fit the data.

\end{itemize}

\section*{Acknowledgements}
MLP acknowledges financial support from the Science and Technology Facilities Council (STFC), and is grateful to Simon Gibbons for helpful discussions. JAT acknowledges partial support from NASA ADAP grant NNX13AE98G.
WNA, EK and ACF acknowledge support from the European Union Seventh Framework Programme (FP7/2013--2017) under grant agreement n.312789, StrongGravity. JW acknowledges support from Deutsches Zentrum
f\"ur Luft- und Raumfahrt grant 50\,OR\,1411.
This work made use of data from the \nustar\ mission, a project led by the California Institute of Technology, managed by the Jet Propulsion Laboratory, and funded by the National Aeronautics and Space Administration. This research has made use of the \nustar\ Data Analysis Software (NuSTARDAS) jointly developed by the ASI Science Data Center (ASDC, Italy) and the California Institute of Technology (USA). This research has made use of data obtained from the \suzaku\ satellite, a collaborative mission between the space agencies of Japan (JAXA) and the USA (NASA). 

\bibliographystyle{mn2e}
\bibliography{bibliography_cygx1}

\end{document}